%% file: benchmark.tex
\pdfoutput=1
\documentclass[11pt]{article}

\usepackage{acl2023}

\usepackage{times}
\usepackage{latexsym}
\usepackage{booktabs}
\usepackage{booktabs}
\usepackage{tabularx}
\usepackage[most]{tcolorbox}
\usepackage[utf8]{inputenc}
\usepackage{amssymb}
\usepackage{amsmath}
\usepackage{textcomp}
\usepackage[T1]{fontenc}
\usepackage[utf8]{inputenc}

\usepackage{microtype}

\usepackage{inconsolata}
\usepackage{amsmath}
\usepackage{latexsym}
\usepackage{graphicx}
\usepackage{enumitem}
\usepackage{xcolor}
\newcommand{\cmark}{\textcolor{green!60!black}{\checkmark}}
\newcommand{\xmark}{\textcolor{red!70!black}{\ding{55}}}
\usepackage{pifont}
\title{CIFE: Code Instruction-Following Evaluation}

\author{
  Sravani Gunnu\thanks{\hspace{0.7em}Work done during an internship at IBM Research India.} \\
  IIT Bombay, India \\
  \texttt{sravanigunnu@cse.iitb.ac.in} 
  \And
  Shanmukha Guttula \\
  IBM Research India \\
  \texttt{shagutt1@in.ibm.com} 
  \And
  Hima Patel \\
  IBM Research India \\
  \texttt{himapatel@in.ibm.com}
}

\begin{document}
\maketitle
\begin{abstract}
Large Language Models (LLMs) are increasingly applied to real-world code generation, where functional correctness alone is insufficient for reliable deployment, developers also expect adherence to explicit requirements for \textit{robustness}, \textit{formatting}, and \textit{security}. Existing benchmarks primarily assess correctness through test-case execution, offering limited insight into how reliably models follow such constraints. We introduce a benchmark of \textbf{1{,}000} Python tasks, each paired with an average of \textbf{7} developer-specified constraints spanning \textbf{13} categories. Constraints are curated through a four-stage human–LLM pipeline to ensure they are \textit{atomic}, \textit{relevant}, and \textit{objective}. We evaluate \textbf{14} open- and closed-source models using complementary adherence metrics and propose the \textbf{C2A Score}, a composite measure that jointly captures correctness and constraint compliance. Results reveal a substantial gap between partial and strict satisfaction, while strong models achieve over \textbf{90\%} partial adherence, strict adherence remains between \textbf{39--66\%}. These findings highlight that trustworthy code generation requires not only correctness but also consistent adherence to developer intent.

\end{abstract}

\begin{figure}[t]
    \centering
    \includegraphics[width=\linewidth]{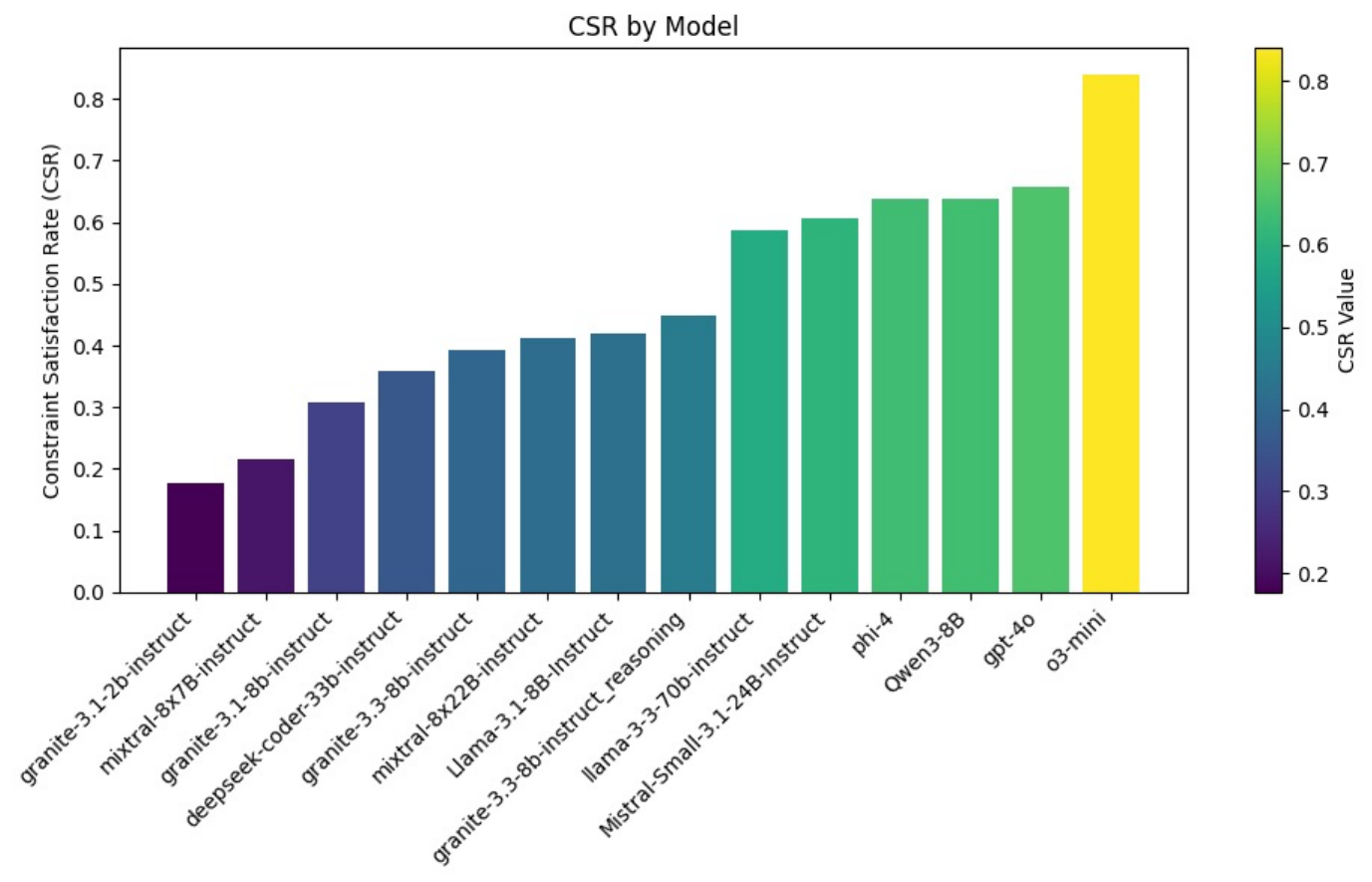}
    \caption{\textbf{Constraint Satisfaction Rate (CSR) across models.} 
    Larger models and those with reasoning capabilities demonstrate significantly higher adherence to developer-specified constraints. 
    The reasoning-oriented \texttt{o3-mini} achieves the highest CSR, while smaller models show steep declines, underscoring the challenge of reliably satisfying multiple coding requirements.}
    \label{fig:csr_by_model}
\end{figure}

\begin{figure*}[h]
    \centering
    \includegraphics[width=\linewidth]{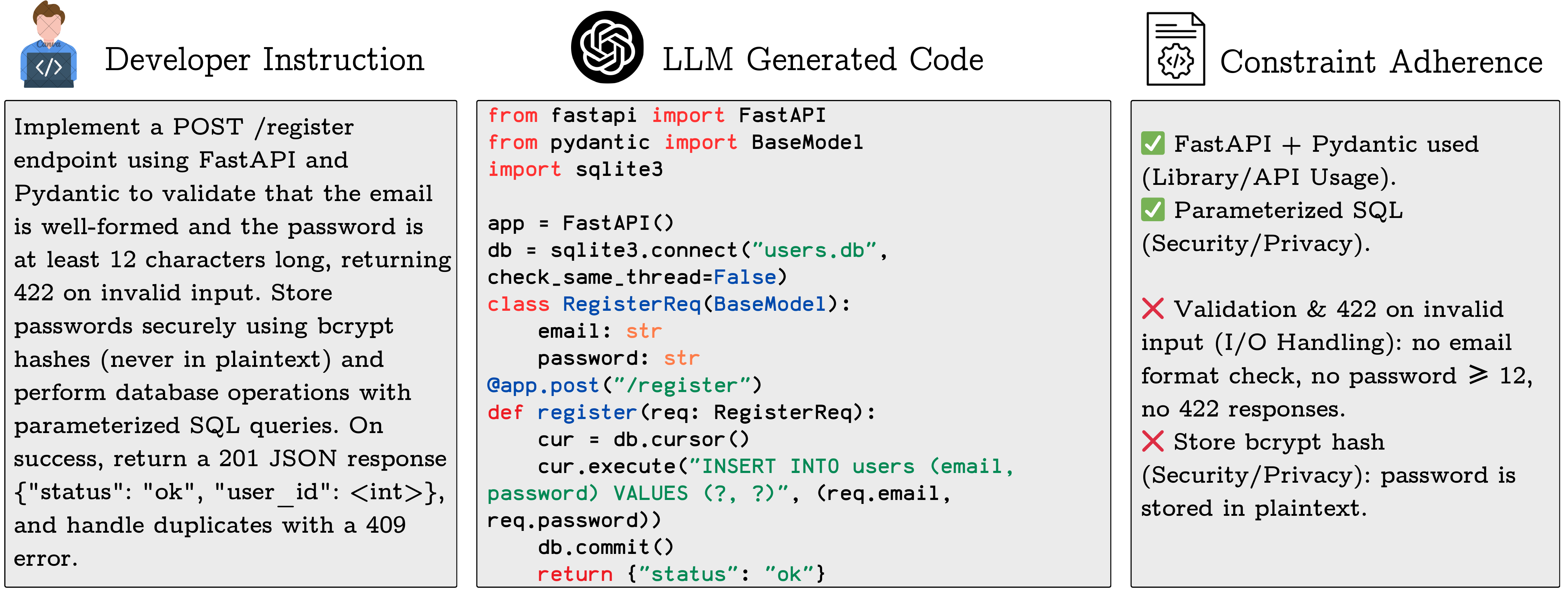}
    \caption{Illustration of constraint adherence in a real-world task. 
    The developer instruction (left) specifies four requirements; the model-generated code (center) satisfies only two. 
    The right panel summarizes satisfied (\cmark ) and violated (\xmark) constraints, showing that syntactically correct code may still fail to meet critical developer requirements.}
    \label{fig:constraint_example}
\end{figure*}

\section{Introduction}

Instruction-following has emerged as a foundational capability for Large Language Models (LLMs), enabling them to understand and execute natural language commands across a broad range of applications, including question answering, summarization, and dialogue systems~\citep{honovich2022unnatural,jiang2023followbench,zheng2023ifeval}.
In the domain of code generation, instruction-following lies at the core of aligning model outputs with developer intent~\citep{codeif2024,codeifbench2024,li-etal-2025-fea,jimenez2024swebench,liu2024fullstackbench}. Beyond generating semantically valid and executable code~\citep{chen2021evaluating,austin2021program,hendrycks2021apps}, a model must accurately interpret and implement the nuanced requirements embedded in developer instructions. These requirements, known as \textit{constraints}, specify how the generated code should behave, encompassing aspects such as input validation, error handling, modularity, and security~\citep{codeif2024,li-etal-2025-fea,liu2024fullstackbench}.
Constraint adherence measures how reliably generated code satisfies these requirements, ensuring outputs are not only correct but also robust, secure, and production-ready~\citep{codeifbench2024,jimenez2024swebench,zhuo2025bigcodebench,liu2024fullstackbench}.

Despite recent advances, LLM-generated code often fails to adhere to critical constraints, leading to incomplete or unreliable implementations~\citep{li2022competition,codeif2024,codeifbench2024,liu2024fullstackbench}. Figure~\ref{fig:constraint_example} illustrates this limitation with a simple web API task, where the model-generated code executes successfully but violates essential constraints such as secure password hashing and input validation.

While instruction-following has been extensively explored in natural language tasks~\citep{honovich2022unnatural,jiang2023followbench,zheng2023ifeval,wen2024complexbench,qin2024infobench,chen2024sifo,efrat2023lmentry,li-etal-2025-fea}, its study within code generation remains comparatively limited. Early benchmarks like HumanEval~\citep{chen2021evaluating}, MBPP~\citep{austin2021program}, and APPS~\citep{hendrycks2021apps} focus on functional correctness (e.g., passing unit tests) but overlook broader developer requirements. Recent benchmarks such as CodeIF~\citep{codeif2024}, CodeIF-Bench~\citep{codeifbench2024}, MultiCodeIF~\citep{multicodeif2025}, and IFEvalCode~\citep{ifevalcode2025} have begun to move beyond correctness-only metrics, but they remain limited, focusing on small-scale or multi-turn setups, relying on narrow or synthetic constraints, and lacking a unified reliability metric. These efforts remain fragmented and they cover parts of the instruction-following challenge but lack dense, developer-grounded constraints and fine-grained single-turn analysis across realistic programming tasks.

To address these limitations, we introduce \textbf{CIFE} (Code Instruction-Following Evaluation), a benchmark that directly targets \textit{constraint adherence} in code generation. CIFE comprises \textbf{1,000 Python programming tasks}, each paired with an average of \textbf{7 constraints} spanning \textbf{13 developer-relevant categories}, and is validated through a systematic pipeline ensuring that the constraints are \textit{atomic}, \textit{relevant}, and \textit{objective}. CIFE fills a critical gap in evaluating how reliably LLMs adhere to developer-specified requirements across realistic coding scenarios, bridging the divide between functional correctness and the broader demands of real-world software development.

\vspace{1mm}
\noindent As instruction-following in code generation becomes a critical area of study, several natural research questions arise:  
\begin{enumerate}[label=\textit{RQ\arabic*.}, leftmargin=*, nosep]
    \item \textit{How does constraint-following ability vary with task complexity?}
    \item \textit{Which constraint types are most difficult (e.g., security, optimization)?}
    \item \textit{Do explicit reasoning capabilities improve constraint adherence?}
    \item \textit{How do model scale and training paradigms affect adherence behavior?}
\end{enumerate}

\vspace{1mm}
\noindent \textbf{Our Contributions.} The key contributions of this paper are as follows:
\begin{enumerate}
[leftmargin=*, nosep]
    \item We introduce \textbf{CIFE} (Code Instruction-Following Evaluation), a benchmark explicitly designed to evaluate \textit{constraint adherence} in Python code generation, capturing diverse developer-specified requirements across 13 categories.
    \item We propose a new composite metric, the \textbf{C2A Score} (Code-Correctness and Constraint-Adherence), which jointly measures code correctness and adherence to developer constraints, enabling a more comprehensive evaluation of LLM behavior.
    \item We conduct a large-scale evaluation of 14 models spanning open-source and proprietary families, varying in size, reasoning capability, and training paradigm. Our results show that reasoning models consistently outperform non-reasoning counterparts,  while adherence tends to decrease as the number of instructions and constraint complexity increase.
    \item We open-source the complete \textbf{CIFE} benchmark, including dataset, evaluation scripts to support further research on instruction-following and constraint adherence.\footnote{\url{https://github.com/IBM/CIFE.git}}
\end{enumerate}

\section{Related Work}

Instruction-following has emerged as a key capability of Large Language Models (LLMs), studied across both natural language and code generation domains. In programming, early benchmarks such as HumanEval~\citep{chen2021evaluating}, MBPP~\citep{austin2021program}, and APPS~\citep{hendrycks2021apps} primarily assessed \textit{functional correctness}, whether generated code passes predefined unit tests—establishing the foundation for automated evaluation but neglecting adherence to nuanced developer requirements. Later efforts extended this paradigm to multilingual and cross-lingual settings~\citep{cassano2022multipl,peng2023humanevalxl,raihan2025mhummaneval}, yet correctness remained the dominant criterion. In contrast, research in natural language instruction-following has emphasized adherence to compositional, constraint-rich, and multi-turn instructions~\citep{zheng2023ifeval,jiang2023followbench,jing2023followeval,wen2024complexbench}, revealing that even advanced LLMs frequently deviate from user-specified conditions when constraints are implicit or multi-dimensional.

Motivated by these findings, recent benchmarks have shifted focus toward evaluating instruction-following in code generation beyond correctness-only metrics. Code-IF~\citep{codeif2024} evaluates static instruction adherence across multiple languages but with a relatively narrow, partially synthetic constraint taxonomy. CodeIF-Bench~\citep{codeifbench2024} explores multi-turn Python interactions but remains small (124 tasks) and emphasizes incremental feedback rather than single-turn reliability. MultiCodeIF~\citep{multicodeif2025} scales to multiple languages and hierarchical instructions but relies heavily on automatically expanded constraints that may not match practical developer concerns. IFEvalCode~\citep{ifevalcode2025} introduces controlled constraint-based evaluation but typically attaches only a few constraints per task and reports correctness and controllability separately, without a unified reliability metric.

Complementary efforts such as BigCodeBench~\citep{bigcodebench2023}, SWE-bench~\citep{jimenez2024swebench}, FEA-bench~\citep{li-etal-2025-fea}, and FullStackBench~\citep{liu2024fullstackbench} extend evaluation to repository-level and end-to-end development scenarios but focus on large-scale or dialogic tasks and do not provide systematic, fine-grained constraint categorization. 

In contrast, \textbf{CIFE} introduces dense, developer-grounded constraint evaluation on realistic Python programming tasks with structured, multi-category constraint evaluation, enabling a comprehensive assessment of how reliably LLMs adhere to developer intent in real-world coding contexts. Our tasks and constraint taxonomy are derived from developer-facing sources and validated through a human–LLM pipeline to ensure each constraint is atomic and practically relevant. CIFE is larger and denser than CodeIF-Bench, offers a broader and more meaningful constraint taxonomy than CodeIF and MultiCodeIF, and contributes a unified reliability metric (C2A Score) combining correctness and strict adherence.

\begin{figure*}[h]
    \centering
    \includegraphics[width=0.85\linewidth]{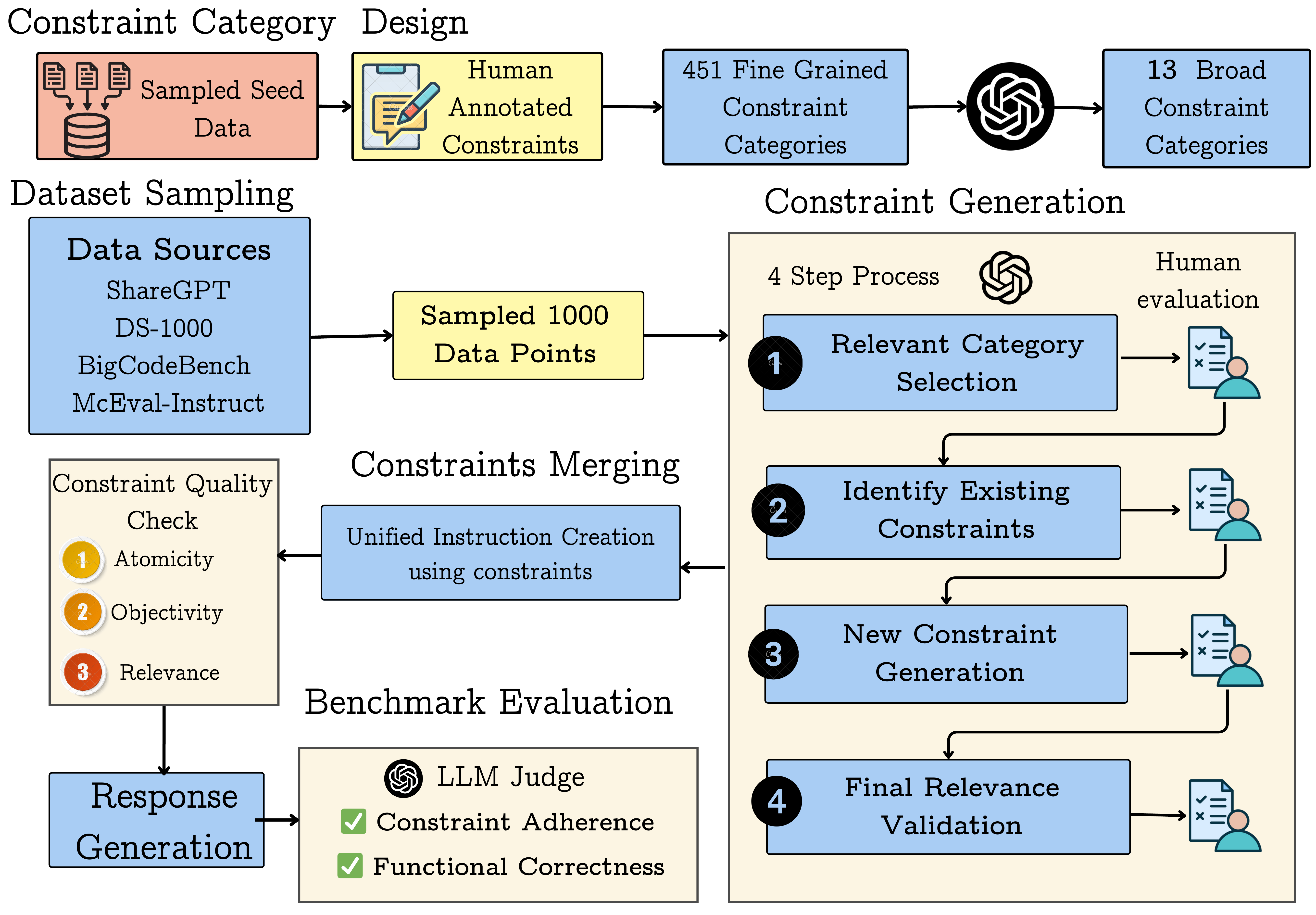}
    \caption{Overview of the benchmark creation workflow, showing task sampling, constraint categorization and generation, quality validation, and final evaluation based on constraint adherence and functional correctness.}
    \label{fig:workflow}
\end{figure*}

\begin{figure*}[h]
    \centering
    \includegraphics[width=\linewidth]{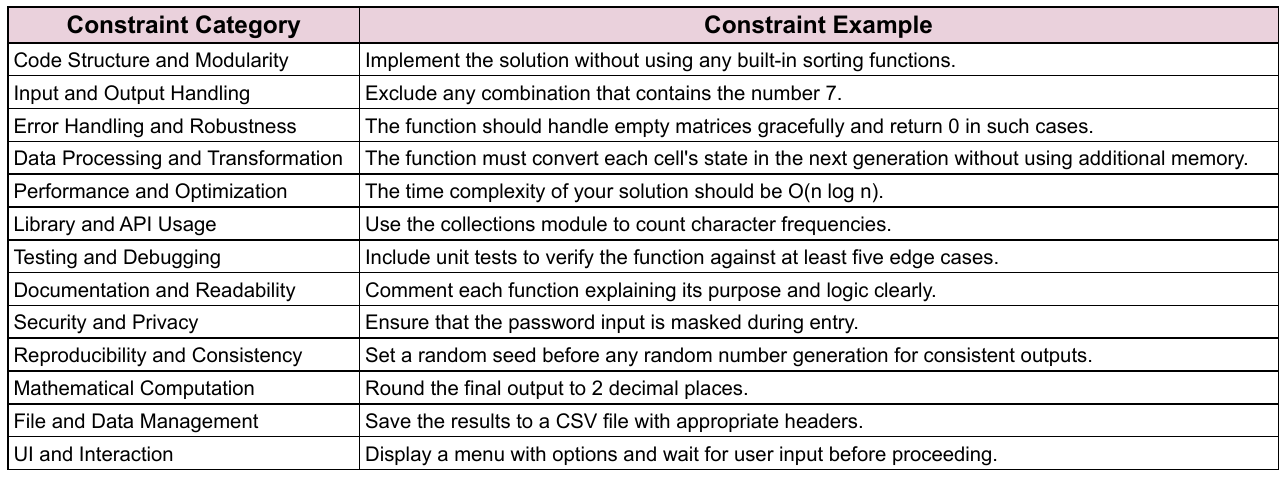}
    \caption{Examples of constraint categories and sample developer-style instructions.}
    \label{fig:constraint_table}
\end{figure*}

\section{Dataset Sources}

To ensure broad coverage of real-world Python programming scenarios, we selected tasks from diverse existing datasets with the help of coding experts. Specifically, we draw from \textit{ShareGPT}~\citep{sharegpt2023}, \textit{DS-1000}~\citep{lai2023ds1000}, \textit{BigCodeBench}~\citep{bigcodebench2023}, and \textit{McEval-Instruct}~\citep{zhong2024mceval}. These sources collectively span conversational, academic, library-intensive, and application-oriented coding problems. From each dataset we selected representative Python tasks and augmented their instructions with explicit developer constraints, creating new instruction–constraint pairs for our benchmark (see Section~\ref{sec:benchmark-workflow}).

\begin{enumerate}
    \item \textbf{ShareGPT} contributes conversationally phrased instructions from real LLM interactions. These tasks capture informal and occasionally ambiguous developer queries, introducing linguistic variation beyond textbook problem statements.

    \item \textbf{DS-1000} provides rigorously validated, test-driven Python exercises. This dataset adds academically grounded tasks with high-quality solutions and precise specifications.

    \item \textbf{BigCodeBench} includes complex tasks involving multiple third-party libraries (e.g., data analysis, visualization, machine learning). These problems represent practical, library-rich development scenarios.

    \item \textbf{McEval-Instruct} offers a diverse set of applied Python tasks emphasizing reasoning, robustness, and realistic developer requirements, complementing the other sources with industry-relevant complexity.
\end{enumerate}

Our final benchmark includes 1,000 tasks: 373 (37.3\%) from McEval-Instruct, 265 (26.5\%) from ShareGPT, 183 (18.3\%) from BigCodeBench, and 179 (17.9\%) from DS-1000. This composition was chosen to balance diversity of sources, real-world task characteristics, and the difficulty of tasks and their associated constraints, yielding a benchmark that is both representative and challenging. Unlike synthetic or contest-style benchmarks, our collection is grounded in real developer contexts and better reflects everyday software development challenges.

\section{Benchmark Creation Workflow}
\label{sec:benchmark-workflow}

Creating a benchmark for instruction-following in code generation requires moving beyond functional correctness to assess whether models can satisfy diverse, fine-grained developer requirements. Unlike existing benchmarks, our goal is to evaluate whether LLMs generate code that adheres to explicit developer-specified constraints spanning diverse aspects such as style, robustness, modularity, security, performance, and other real-world software requirements.

\noindent \textbf{Definition of Constraint:}
A \textit{constraint} is an explicit developer-specified requirement that governs how code should be written or behave, encompassing aspects such as robustness, modularity, formatting consistency, and performance efficiency. An example of such constraints is provided in Appendix~\ref{app:merge-instruction}.

Figure~\ref{fig:workflow} presents an overview of our benchmark creation and evaluation pipeline, which comprises four major stages: (1) \textit{constraint category design}, (2) \textit{constraint generation}, (3) \textit{constraint merging}, and (4) \textit{evaluation methodology}. This structured framework ensures that constraints are systematically designed, rigorously validated, coherently integrated into task descriptions, and consistently evaluated to capture fine-grained instruction-following behavior in realistic coding scenarios.

\subsection{Constraint Category Design}

\paragraph{Constraint Taxonomy.}
As illustrated in Figure~\ref{fig:workflow}, benchmark construction begins with designing a taxonomy of developer-relevant constraints. We first annotated sampled tasks from the source datasets for constraint patterns and then used LLM-assisted clustering to derive fine-grained characteristics, which were consolidated into \textbf{13 broad categories}. These categories capture common real-world requirements such as \textit{Error Handling and Robustness}, \textit{Performance and Optimization}, and \textit{Security and Privacy}. Figure~\ref{fig:constraint_table} presents the constraint categories and corresponding developer-style examples, illustrating the diversity and practical scope of the taxonomy.

\subsection{Constraint Generation Pipeline.}
Each programming task in CIFE is paired with a corresponding set of constraints generated through a structured four-stage pipeline, as illustrated in Figure~\ref{fig:workflow}. At each stage, the \textbf{GPT-4o-mini} model was employed to systematically identify, generate, and validate constraints, ensuring comprehensive coverage of real-world developer requirements. A complete example demonstrating all four stages is provided in Appendix~\ref{app:CG_example}, and the prompts used in each stage are listed in Appendix~\ref{app:prompts}.

The first stage, \textbf{Relevant Category Selection}, identifies which of the 13 high-level constraint categories apply to a given instruction, ensuring contextual grounding by narrowing the scope to only relevant types such as \textit{Error Handling and Robustness} or \textit{Security and Privacy}. The resulting selected categories, as shown in  Figure~\ref{fig: stage_1_op}, guide subsequent stages.

The second stage, \textbf{Identifying Existing Constraints}, focuses on identifying explicit requirements already present in the instruction. For instance, if the instruction states that “results must be returned in JSON format,” this requirement is preserved under the \textit{Input and Output Handling} category. As shown in Figure~\ref{fig: stage_2_op}, this step ensures that all such explicit constraints are systematically captured for subsequent evaluation.

In the third stage, \textbf{New Constraint Generation}, the LLM (\textit{GPT-4o-mini}) generates additional semantically aligned constraints based on the task and categories selected in Stage 1. These constraints are designed to be realistic and contextually consistent with the task, expanding coverage with developer-relevant expectations such as “return appropriate HTTP status codes” or “implement input validation to prevent injection attacks.” An illustration is shown in Figure~\ref{fig: stage_3_op}.

Finally, the fourth stage, \textbf{Final Relevance Validation}, refines the combined set of constraints by removing redundant, vague, or subjective ones. Constraints that are atomic, relevant, and objectively verifiable are retained. Both human annotators and LLM validators contribute to this process to ensure quality and consistency, producing the final benchmark-ready set shown in Figure~\ref{fig: stage_4_op}.

This four-stage process ensures that each task in CIFE is paired with a high-quality, diverse, and validated set of constraints aligned with developer intent.

\begin{figure}[h!]
    \centering
    \includegraphics[width=\linewidth]{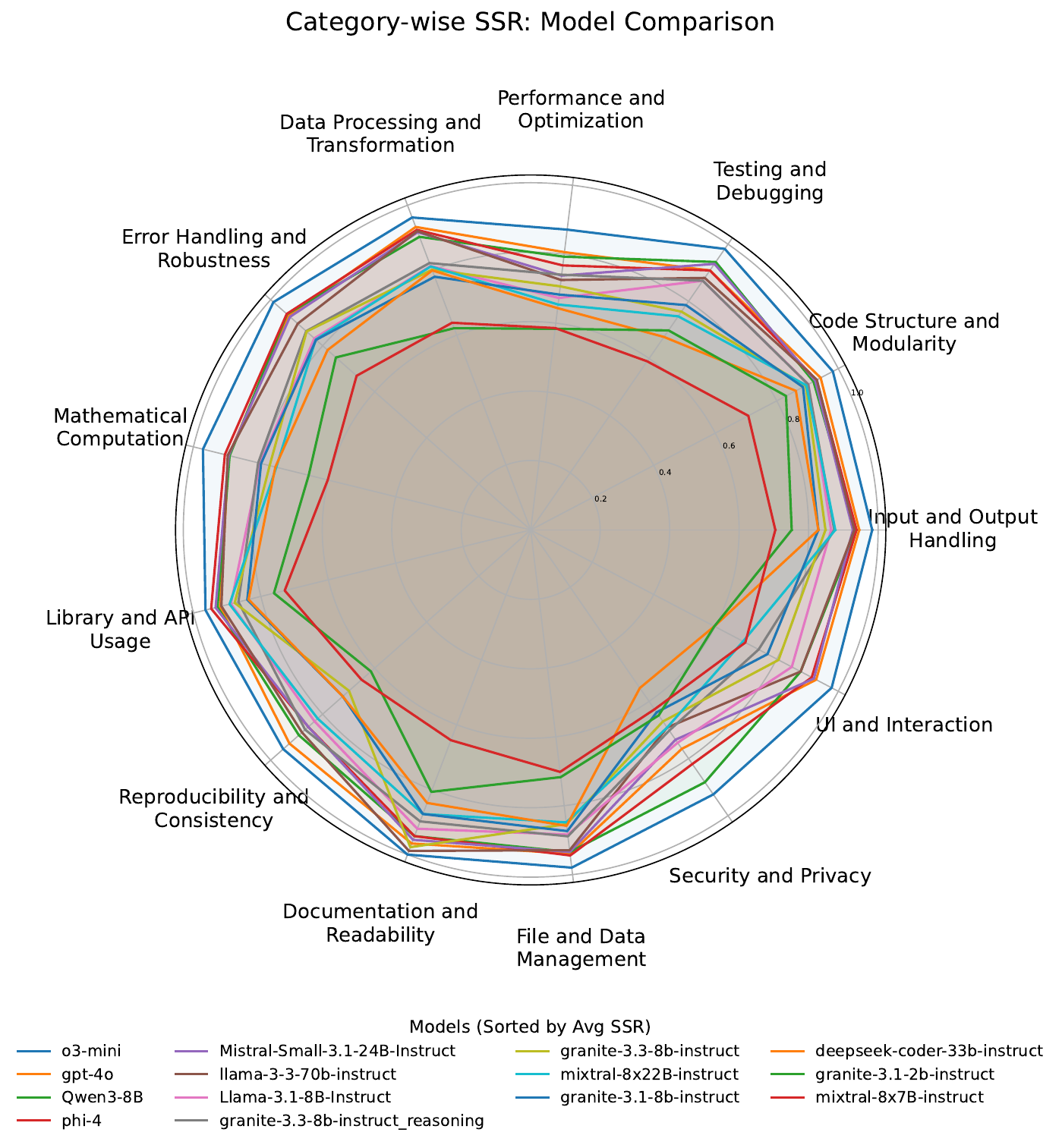}
    \caption{Category-wise SSR comparison across models. Constraints related to security/privacy and optimization are consistently the hardest to follow.}
    \label{fig:category_ssr}
\end{figure}

\subsection{Constraint Merging}
Once the constraints were generated, they were integrated back into the original task descriptions to form unified evaluation prompts. Each constraint was inserted at contextually appropriate points with minimal paraphrasing, preserving the semantics and structure of the instruction. The prompt used and an illustrative example are provided in Figure~\ref{fig:merge_prompt} and Figure~\ref{fig:merge_example}, respectively.

\subsection{Evaluation Methodology}

We adopt an \textit{LLM-as-Judge} framework~\citep{zheng2023mtbench} to assess model outputs. In this setup, a strong reference model, \textbf{GPT-4o-mini}, evaluates responses according to structured rubrics along two complementary dimensions: \textit{constraint adherence} and \textit{functional correctness}. This dual evaluation captures whether the generated code aligns with developer-specified requirements while also fulfilling the intended functionality.

\paragraph{Constraint Adherence.}
For each task, the LLM determines whether every constraint is satisfied (\texttt{true}) or violated (\texttt{false}), producing an adherence vector per example from which strict and soft adherence metrics are derived~\citep{codeif2024,codeifbench2024}. The evaluation prompt is shown in Figure~\ref{fig:llm_judge_constraint}. To ensure reliability, human annotators independently validated a subset of judgments, and agreement between human and LLM evaluations is reported in Appendix~\ref{app:human-annotation}.

\paragraph{Functional Correctness.}
Traditional evaluations using predefined unit tests or static outputs~\citep{chen2021evaluating} face scalability and coverage limitations in open-ended instruction-following tasks. They rely on handcrafted input–output pairs and cannot effectively assess reasoning-based or non-executable tasks, where correctness is semantic rather than behavioral~\citep{li2022competition}. To address this, we employ the \textit{LLM-as-Judge} framework~\citep{zheng2023mtbench}, which enables scalable and semantically consistent evaluation. The evaluator assigns one of three labels \textbf{completely correct}, \textbf{partially correct}, or \textbf{incorrect} to each model response, as detailed in Figure~\ref{fig:llm_judge_correctness}.

\paragraph{Evaluation Metrics.}  
We compute three metrics to capture different aspects of instruction-following ability:  

\begin{enumerate}
    \item \textbf{Constraint Satisfaction Rate (CSR).}  
    CSR measures strict adherence by checking whether \textit{all} constraints associated with a task are satisfied. Formally, for task $i$ with $k_i$ constraints and adherence vector $\mathbf{a}_i \in \{0,1\}^{k_i}$
    \[
    \text{CSR} = \frac{1}{N} \sum_{i=1}^{N} \mathbf{1}\!\left[ \sum_{j=1}^{k_i} a_{ij} = k_i \right],
    \]
    where $N$ is the total number of tasks.  

    \item \textbf{Soft Satisfaction Rate (SSR).}  
    SSR captures partial adherence by averaging the proportion of constraints satisfied per task
    \[
    \text{SSR} = \frac{1}{N} \sum_{i=1}^{N} \frac{1}{k_i} \sum_{j=1}^{k_i} a_{ij}
    \]
    This metric rewards models that satisfy a majority of constraints, even if not all are met.  

    \item \textbf{C2A Score (Code-Correctness and Constraint-Adherence).}  
    To jointly evaluate correctness and constraint-following, we introduce the \textbf{C2A Score}, a new composite metric. It measures the fraction of model responses that are both functionally correct and satisfy all constraints
    \[
    \text{C2A} = \frac{1}{N} \sum_{i=1}^{N} \mathbf{1}\!\big[ \text{Correct}(y_i) \wedge \big(\sum_{j=1}^{k_i} a_{ij} = k_i \big) \big],
    \]
    where $\text{Correct}(y_i)$ indicates that the model output for task $i$ is judged as completely correct by the LLM judge.  
\end{enumerate}

While CSR and SSR separately capture strict and soft adherence to constraints, \textbf{C2A is the first metric to jointly measure correctness and constraint adherence}, providing a more holistic view of instruction-following in code generation.

\begin{figure}[h!]
    \centering
    \includegraphics[width=0.8\linewidth]{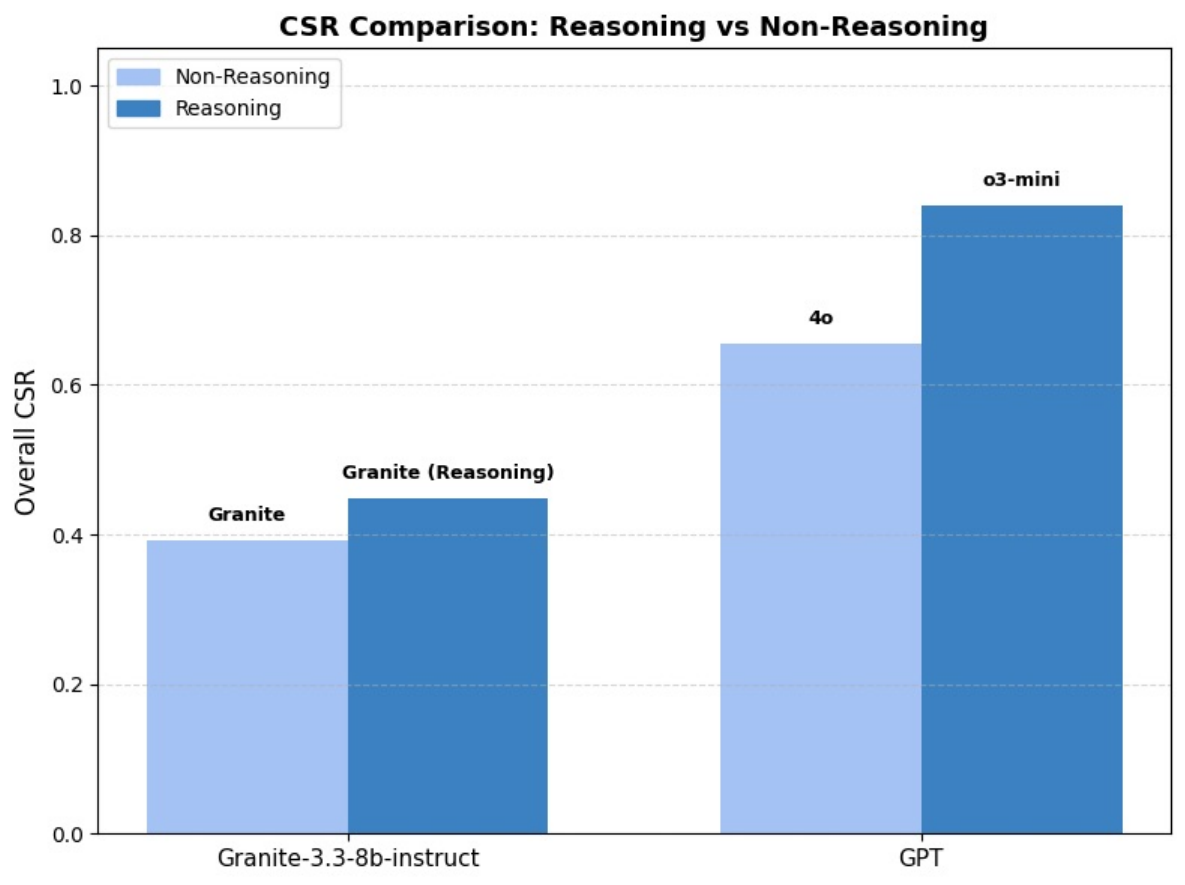}
    \caption{CSR comparison of reasoning vs non-reasoning variants. Explicit reasoning capabilities improve adherence, e.g., Granite-Reasoning vs Granite.}
    \label{fig:reasoning_vs_nonreasoning}
\end{figure}

\section{Constraint Quality Validation}
In order to access the quality of the generated constraints, we validated them to ensure their reliability for benchmarking. A sample of data points was manually annotated by human experts, who rated each constraint on a \textbf{scale of 1 to 5} across three dimensions \textbf{atomicity}, \textbf{relevance}, and \textbf{objectivity} achieving mean scores of \textbf{4.87}, \textbf{4.92}, and \textbf{4.82}, respectively. To scale validation across the full dataset, an LLM-based evaluator (\textit{GPT-4o-mini}) was employed using the structured prompt described in Figure ~\ref{fig:quality_prompt} in Appendix, yielding an \textbf{85\%} agreement with human annotations. Further details of the agreement analysis are provided in Appendix~\ref{app:human-annotation}.
\section{Experimental Setup}

We evaluated 14 large language models spanning a range of architectures, sizes, and training paradigms. The Granite family includes \textbf{Granite-3.1-2B-Instruct} and \textbf{Granite-3.1-8B-Instruct} as earlier baselines, together with \textbf{Granite-3.3-8B-Instruct} and its reasoning-enhanced variant \textbf{Granite-3.3-8B-Instruct Reasoning}. From the Mistral family, we include both mixture-of-experts and dense transformer models: \textbf{Mixtral-8x7B-Instruct-v0.1}, \textbf{Mixtral-8x22B-Instruct-v0.1}, and \textbf{Mistral-Small-3.1-24B-Instruct-2503}. The LLaMA family is represented by \textbf{LLaMA-3.1-8B-Instruct} and \textbf{LLaMA-3-70B-Instruct}, while we also evaluate \textbf{DeepSeek-Coder-33B-Instruct}, optimized for code generation tasks. In addition, we consider Microsoft’s \textbf{Phi-4} and Alibaba’s \textbf{Qwen3-8B}, both strong open-source instruction-tuned models. Finally, we include two proprietary models \textbf{GPT-4o}, a large non-reasoning model, and \textbf{O3-Mini}, a smaller reasoning-oriented model that achieved the strongest overall results in our evaluation. All models were prompted with the unified instruction containing both the original programming task and associated constraints, and decoding was performed with temperature fixed at 0 to minimize randomness and ensure comparability.
All models were prompted with unified instructions containing both the original programming task and its associated constraints. Response generation used the maximum context length supported by each model as the 
\texttt{max\_tokens} limit, with decoding temperature fixed at 0 to ensure deterministic outputs. For the \textit{LLM-as-Judge} evaluations, we used \textbf{GPT-4o-mini} with a temperature of 0.1 and a \texttt{max\_tokens} limit of 2048 to balance diversity and evaluation completeness.

% --- Color defs (put in preamble once) ---

% preamble: \usepackage{booktabs} \usepackage{tabularx}
% and once:  \newcolumntype{Y}{>{\raggedright\arraybackslash}X}

% \begin{table}[t]
% \centering
% \small
% \setlength{\tabcolsep}{6pt}
% \begin{tabularx}{\linewidth}{Y c c c}
% \toprule
% \textbf{Models} & \textbf{CSR} & \textbf{SSR} & \textbf{C2A} \\
% \midrule
% Granite-3.1-2B-Instruct             & 0.177 & 0.6913 & 0.055 \\
% Mixtral-8x7B-Instruct-v0.1          & 0.215 & 0.6581 & 0.052 \\
% Granite-3.1-8B-Instruct             & 0.309 & 0.8008 & 0.117 \\
% DeepSeek-Coder-33B-Instruct         & 0.358 & 0.7941 & 0.172 \\
% Granite-3.3-8B-Instruct             & 0.393 & 0.8237 & 0.152 \\
% Mixtral-8x22B-Instruct-v0.1         & 0.412 & 0.8277 & 0.170 \\
% LLaMA-3.1-8B-Instruct               & 0.419 & 0.8318 & 0.166 \\
% Granite-3.3-8B-Instruct\_Reasoning  & 0.449 & 0.8443 & 0.200 \\
% LLaMA-3-3-70B-Instruct              & 0.586 & 0.9039 & 0.305 \\
% Mistral-Small-3.1-24B-Instruct-2503 & 0.607 & 0.9083 & 0.355 \\
% Phi-4                                & 0.637 & 0.9153 & 0.382 \\
% Qwen3-8B                             & 0.638 & 0.9137 & 0.310 \\
% GPT-4o-2024-08-06                    & 0.656 & 0.9242 & 0.479 \\
% O3-Mini                              & \textbf{0.840} & \textbf{0.9670} & \textbf{0.720} \\
% \bottomrule
% \end{tabularx}
% \caption{Overall results on CIFE. CSR = strict constraint adherence, SSR = soft adherence, and C2A = joint correctness+adherence. Best per column in bold.}
% \label{tab:overall_results_clean}
% \end{table}

% Preamble (once)

\begin{table}[t]
\centering
\small
\setlength{\tabcolsep}{3.5 pt} % optional
\begin{tabular*}{\linewidth}{@{\extracolsep{\fill}} l c c c @{}}
\toprule
\textbf{Models} & \textbf{CSR} & \textbf{SSR} & \textbf{C2A} \\
\midrule
Granite-3.1-2B-Instruct             & 0.177 & 0.6913 & 0.055 \\
Mixtral-8x7B-Instruct          & 0.215 & 0.6581 & 0.052 \\
Granite-3.1-8B-Instruct             & 0.309 & 0.8008 & 0.117 \\
DeepSeek-Coder-33B-Instruct         & 0.358 & 0.7941 & 0.172 \\
Granite-3.3-8B-Instruct             & 0.393 & 0.8237 & 0.152 \\
Mixtral-8x22B-Instruct         & 0.412 & 0.8277 & 0.170 \\
LLaMA-3.1-8B-Instruct               & 0.419 & 0.8318 & 0.166 \\
Granite-3.3-8B-Instruct\_Reasoning  & 0.449 & 0.8443 & 0.200 \\
LLaMA-3-3-70B-Instruct              & 0.586 & 0.9039 & 0.305 \\
Mistral-Small-3.1-24B-Instruct & 0.607 & 0.9083 & 0.355 \\
Phi-4                                & 0.637 & 0.9153 & 0.382 \\
Qwen3-8B                             & 0.638 & 0.9137 & 0.310 \\
GPT-4o-2024-08-06                    & 0.656 & 0.9242 & 0.479 \\
O3-Mini                              & \textbf{0.840} & \textbf{0.9670} & \textbf{0.720} \\
\bottomrule
\end{tabular*}
\caption{Overall CIFE results. CSR denotes strict constraint adherence, SSR denotes soft adherence, and C2A denotes joint correctness and adherence.}

\label{tab:overall_results}
\end{table}

\section{Results}

We evaluate 14 models spanning different families and parameter scales, and the results reveal several consistent trends. First, as shown in Table~\ref{tab:overall_results}, there is a large gap between soft and strict adherence: while most models achieve SSR above 0.80, strict CSR remains much lower, ranging from 0.17 for Granite-3.1-2B to 0.84 for \texttt{gpt-o3-mini}. Our proposed C2A score, which combines correctness and constraint satisfaction, further underscores this difficulty, with even the best-performing \texttt{gpt-o3-mini} reaching only 0.72. 

\begin{figure}[h!]
    \centering
    \includegraphics[width=\linewidth]{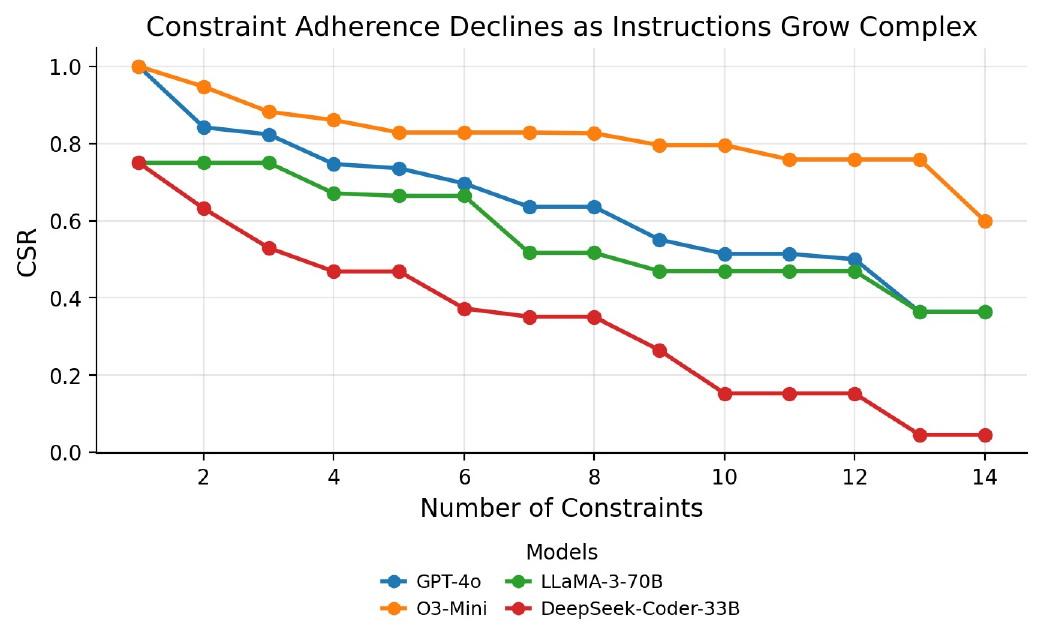}
    \caption{Constraint adherence (CSR) declines as the number of developer-specified requirements increases. High-performing models such as O3-Mini sustain adherence longer, whereas weaker models like DeepSeekCoder-33B drop sharply. This trend highlights the difficulty of satisfying multiple constraints in real-world coding tasks.}
    \label{fig:csr_vs_constraints_intro}
\end{figure}

Second, constraint adherence declines sharply as the number of requirements grows, as illustrated in Figure~\ref{fig:csr_vs_constraints_intro}. Although \texttt{gpt-o3-mini} sustains higher CSR across increasing constraint counts, both GPT-4o and LLaMA-3-70B show a downward trend, and weaker models such as DeepSeek-Coder-33B drop steeply, highlighting the compositional challenge of satisfying multiple constraints simultaneously.  

Third, category-level analysis in Figure~\ref{fig:category_ssr} shows that while models perform relatively well on input/output handling and error robustness, they consistently struggle with security/privacy and optimization-related requirements, indicating that these developer concerns remain particularly challenging. Figure~\ref{fig:csr_by_difficulty} further breaks down performance by task difficulty, showing that all models degrade from easy to hard tasks, but reasoning-enhanced models exhibit more graceful declines.  

Finally, Figure~\ref{fig:reasoning_vs_nonreasoning} directly compares reasoning and non-reasoning variants, revealing measurable improvements when explicit reasoning capabilities are present. Within the Granite family, the reasoning model (\textit{granite-3.3-8b-instruct\_reasoning}) consistently outperforms its non-reasoning counterpart, underscoring the benefits of reasoning alignment. In the GPT family, a particularly notable insight is that \texttt{gpt-o3-mini} (reasoning, $\sim$3.8B parameters) surpasses GPT-4o (non-reasoning, $\sim$175B parameters) by a significant margin in CSR, highlighting that reasoning ability can outweigh sheer model scale in constraint-following tasks.  

Collectively, these results demonstrate that while state-of-the-art models have made progress in partial adherence, reliably satisfying complex, multi-dimensional developer instructions remains an open challenge.

\section{Ablation and Analysis}
Our analyses allow us to revisit the key questions posed in the introduction. First, constraint-following ability clearly varies with task complexity: Figure~\ref{fig:csr_by_difficulty} shows that CSR drops substantially from easy to hard tasks across all models. Second, some constraint types are demonstrably harder than others, with Figure~\ref{fig:category_ssr} highlighting that security/privacy and optimization requirements remain the most challenging, whereas input/output handling and error robustness are comparatively easier. Third, reasoning capabilities provide consistent gains: reasoning-enhanced variants such as \texttt{granite-3.3-8b-instruct\_reasoning} and \texttt{gpt-o3-mini} outperform their non-reasoning counterparts, as shown in Figure~\ref{fig:reasoning_vs_nonreasoning}. Fourth, model scale alone is not a reliable predictor of adherence. Notably, \texttt{gpt-o3-mini} ($\sim$3.8B) surpasses GPT-4o ($\sim$175B) in CSR, underscoring that reasoning alignment can outweigh parameter count.

Our experiments over 14 models reveal consistent patterns: (i) a large gap between soft and strict adherence, (ii) sharply declining adherence as the number and difficulty of constraints increase, and (iii) systematic gains from explicit reasoning often outweighing raw parameter scale. Category-level analyses further show persistent weaknesses in optimization and security/privacy requirements. These findings collectively highlight that constraint adherence is shaped by a combination of task difficulty, constraint type, reasoning ability, and training paradigm, rather than scale alone.

\section{Conclusion and Future Work}

We introduced \textbf{CIFE}, a benchmark for evaluating \emph{constraint adherence} in code generation, comprising 1,000 Python tasks with fine-grained, validated constraints across developer-relevant categories. Beyond correctness, CIFE measures whether LLMs meet explicit requirements via \emph{CSR} and \emph{SSR} and offers \emph{C2A}, a composite metric that jointly assesses correctness and full adherence. Future work includes extending CIFE beyond Python to multi-language and repository-level settings, enabling interactive multi-turn workflows with tool use, exploring hybrid evaluation that combines LLM judging with static and dynamic analyses, supporting richer verifiable constraints such as security policies and performance budgets, and improving training strategies for constraint-following models.

\section*{Limitations}
CIFE currently targets Python and single-turn generation with an LLM-as-judge protocol. These choices may limit generalizability. For example, our method is untested on multi-language or repository-level tasks, interactive multi-turn workflows, or hybrid evaluation settings.

\bibliographystyle{acl_natbib}
\bibliography{benchmark_references,custom,anthology} % include whichever .bib files contain your keys
\appendix
\clearpage
\input{appendix}

\end{document}

%% file: appendix.tex
% =========================
% appendix.tex
% =========================
\appendix
\section*{Appendix}
\addcontentsline{toc}{section}{Appendix}

% ---- Packages needed (put in preamble if not already present) ----
% \usepackage[most]{tcolorbox}
% \usepackage{listings}
% \usepackage{xcolor}

% Optional, consistent code/JSON look
\lstdefinestyle{cifejson}{
  basicstyle=\ttfamily\small,
  columns=fullflexible,
  breaklines=true,
  frame=single,
  framerule=0pt,
  backgroundcolor=\color{gray!6},
  showstringspaces=false,
}

\tcbset{
  cife/.style={
    enhanced,
    breakable,
    colback=gray!2,
    colframe=black!8,
    coltitle=black,
    boxrule=0.4pt,
    arc=2pt,
    left=6pt,right=6pt,top=6pt,bottom=6pt
  },
  cifeheader/.style={
    enhanced,
    boxrule=0pt,
    colback=gray!2,
    fonttitle=\bfseries,
    arc=2pt,
    left=6pt,right=6pt,top=4pt,bottom=4pt
  }
}

\section{Example of the Constraint-Generation Pipeline}
\label{app:CG_example}
This appendix illustrates the four stages used to construct constraints for a single instruction. At each stage, the output is generated by a dedicated prompt (see \S\ref{app:prompts} in the appendix for the exact prompts).

% ------------------------
% Instruction Box
% ------------------------
\begin{figure}[h!]
\centering
\begin{tcolorbox}[title=Instruction, colback=gray!5!white, colframe=black!75!white, fontupper=\ttfamily\small]

Write a Python Flask application that provides a REST API to manage a simple inventory of products. Each product should have an `id`, `name`, and `quantity` as its attributes. The API should allow clients to perform the following operations:
\begin{enumerate}[nosep]
    \item List all products
    \item Get a single product by its `id`
    \item  Create a product
    \item Update a product
    \item Delete a product by its `id`
\end{enumerate}

Use Flask for the web framework and Flask Marshmallow for serialization/deserialization of product data. Ensure the application handles cases where a product with a given `id` does not exist.
\end{tcolorbox}
\caption{Example Instruction}
\label{fig: instruction}
\end{figure}
% ------------------------
% Stage 1
% ------------------------

\begin{figure}[h!]
\centering
\begin{tcolorbox}[title=Stage 1: Relevant Category Selection, colback=gray!5!white, colframe=black!75!white, fontupper=\ttfamily\small]

\textbf{Goal.} Select which of the benchmark’s high-level categories are applicable to the instruction.\vspace{2mm}

\textbf{Output (categories selected).}
\begin{enumerate}
[nosep]
  \item Code Structure and Modularity
  \item Input and Output Handling
  \item Error Handling and Robustness
  \item Data Processing and Transformation
  \item Library and API Usage
  \item Testing and Debugging
  \item Documentation and Readability
  \item Security and Privacy
  \item Reproducibility and Consistency
\end{enumerate}

\end{tcolorbox}
\caption{Selecting relevant categories}
\label{fig: stage_1_op}
\end{figure}

% ------------------------
% Stage 2
% ------------------------
\begin{figure}[h!]
\centering
\begin{tcolorbox}[title=Stage 2: Identifying Existing Constraints, colback=gray!5!white, colframe=black!75!white, fontupper=\ttfamily\small]

\textbf{Goal: } Extract explicit requirements already present in the instruction and align each with a category.\vspace{2mm}

\textbf{Existing Constraints:}
\begin{enumerate}
 
  \item \textbf{Library and API Usage:} Use Flask for the web framework.
  \item \textbf{Library and API Usage:} Use Flask Marshmallow for serialization or deserialization of product data.
  \item \textbf{Error Handling and Robustness:} Handle cases where a product with a given \texttt{id} does not exist.
  \item \textbf{Input and Output Handling:} Support list, get, create, update, and delete operations for products.

\end{enumerate}
\end{tcolorbox}
\caption{Identifying existing constraints}
\label{fig: stage_2_op}
\end{figure}

% ------------------------
% Stage 3
% ------------------------
\begin{figure}[h!]
\centering
\begin{tcolorbox}[title=Stage 3: New Constraint Generation, colback=gray!5!white, colframe=black!75!white, fontupper=\ttfamily\small]
\textbf{Goal.} Propose additional \emph{semantically aligned} constraints within the selected categories to reflect realistic developer expectations.\vspace{2mm}

\textbf{Output (candidate constraints).}
\begin{itemize}[leftmargin=1.5em]
  \item \textbf{Input and Output Handling:} The API must return a JSON response for all endpoints.
  \item \textbf{Input and Output Handling:} Accept JSON input for creating and updating products.
  \item \textbf{Error Handling and Robustness:} Return appropriate HTTP status codes for each operation (e.g., 200 success; 404 not found; 201 created).
  \item \textbf{Code Structure and Modularity:} Organize the code into functions for each endpoint to enhance readability and maintainability.
  \item \textbf{Testing and Debugging:} Include unit tests for each endpoint.
  \item \textbf{Documentation and Readability:} Provide clear comments/docstrings for each function and endpoint.
  \item \textbf{Security and Privacy:} Implement input validation to prevent injection attacks and ensure data integrity.
\end{itemize}

\end{tcolorbox}
\caption{Generating new constraints}
\label{fig: stage_3_op}
\end{figure}

% ------------------------
% Stage 4
% ------------------------
\begin{figure}[h!]
\centering
\begin{tcolorbox}[title=Stage 4: Final Relevance Validation, colback=gray!5!white, colframe=black!75!white, fontupper=\ttfamily\small]
\textbf{Goal.} Filter for \emph{atomic}, \emph{relevant}, and \emph{objectively verifiable} constraints; remove redundancies and vague items.\vspace{2mm}

\textbf{Finalized Constraints (kept).}
\begin{itemize}[leftmargin=1.5em]
  \item \textbf{Library and API Usage:} Use Flask for the web framework.
  \item \textbf{Library and API Usage:} Use Flask–Marshmallow for serialization or deserialization.
  \item \textbf{Error Handling and Robustness:} Handle missing-product \texttt{id} gracefully.
  \item \textbf{Input and Output Handling:} Return JSON responses for all endpoints.
  \item \textbf{Input and Output Handling:} Accept JSON input for create/update operations.
  \item \textbf{Error Handling and Robustness:} Use appropriate HTTP status codes (200/201/404).
  \item \textbf{Security and Privacy:} Validate inputs to prevent injection and preserve data integrity.
\end{itemize}

\textbf{Removed (non-atomic or out-of-scope).}
\begin{itemize}[leftmargin=1.5em]
  \item \textbf{Code Structure and Modularity:} Organize endpoints into functions (kept implicitly by coding style; removed as not directly verifiable in some solutions).
  \item \textbf{Testing and Debugging:} Include unit tests (excluded for tasks lacking test harnesses).
  \item \textbf{Documentation and Readability:} Provide comments/docstrings (removed when not objectively checkable for all tasks).
\end{itemize}

\end{tcolorbox}
\caption{Validating the generated constraints}
\label{fig: stage_4_op}
\end{figure}

\section{Prompt Templates}
\label{app:prompts}
For completeness, the exact prompts used at each stage are provided here.

\paragraph{P1: Relevant Category Selection.} 
The prompt for selecting relevant constraint categories is shown in Figure~\ref{fig:stage1_prompt}, which guides the model to identify all applicable high-level categories from the benchmark taxonomy based on the given instruction and code.

\begin{figure*}[h!]
\centering
\begin{tcolorbox}[title=Prompt 1: Relevant Category Selection , colback=gray!5!white, colframe=black!75!white, fontupper=\ttfamily\small]

\textbf{Goal:} Classify the natural language instruction and its corresponding code into all relevant high-level constraint categories from a predefined comprehensive list.

\textbf{Context:} You are given an instruction describing a coding task, its corresponding code for reference, and a list of all possible constraint categories.

\textbf{Task Description:} Select \textbf{all} categories that apply directly, indirectly, or potentially to the given instruction and code.  
Consider both explicit requirements (e.g., mentioned constraints) and implicit expectations (e.g., robustness, structure, or efficiency).  
If there is any plausible reason for a category to be relevant — due to instruction wording, code structure, or real-world use — include it.  
The goal is to maximize coverage, ensuring no relevant category is missed.
\textbf{Inputs Required}
\begin{verbatim}
 {
        "instruction": {instruction}
        "code": {code}
        "all_constraint_categories": {categories_str}
    }
\end{verbatim}
\textbf{Output Format:}
\begin{verbatim}
{
  "relevant_categories": [
    "List of selected relevant categories from all_constraint_categories"
  ]
}
\end{verbatim}
\end{tcolorbox}
\caption{Prompt used in Stage 1: Relevant Category Selection.}
\label{fig:stage1_prompt}

\end{figure*}

\paragraph{P2: Identifying Existing Constraints.}
The prompt for identifying explicit constraints embedded within the instruction is shown in Figure~\ref{fig:stage2_prompt}. It guides the model to extract all atomic, explicit constraints while preserving the core problem statement.

\begin{figure*}[h!]
\centering
\begin{tcolorbox}[title=Prompt 2: Identifying Existing Constraints, colback=gray!5!white, colframe=black!75!white, fontupper=\ttfamily\small]

\textbf{Goal:} Carefully analyze the provided programming instruction to separate the core problem description from any embedded explicit constraints or directives.

\textbf{Context:} You are given an instruction describing a coding task and a list of relevant constraint categories.  
The instruction may contain specific rules, formatting requirements, or implementation details that must be extracted.

\textbf{Task Description:}  
\begin{itemize}[leftmargin=1.5em, nosep]
    \item Identify and extract all \textbf{explicit constraints} within the instruction, such as required function names, variable handling, algorithms, documentation expectations, or formatting rules.  
    \item Split compound constraints into \textbf{atomic} ones, ensuring that each extracted constraint refers to exactly one condition or requirement.   
    \item Format extracted constraints as JSON objects, each labeled with an appropriate category from the provided list.  
\end{itemize}

\textbf{Example of Atomic Extraction:}
\begin{verbatim}
Original: "Raise ValueError if input is None or if it's an empty list."
Extracted:
{
  "type": "Error Handling and Robustness",
  "constraint": "Raise ValueError if input is None.",
},
{
  "type": "Error Handling and Robustness",
  "constraint": "Raise ValueError if input is an empty list.",
 
}
\end{verbatim}

\textbf{Output Format:}
\begin{lstlisting}[style=cifejson]
{
  "extracted_constraints": [
    {
      "type": "Constraint_Category_Name",
      "constraint": "A single, atomic extracted constraint.",
    }
  ]
}
\end{lstlisting}
\end{tcolorbox}
\caption{Prompt used in Stage 2: Identifying Existing Constraints.}
\label{fig:stage2_prompt}
\end{figure*}

\paragraph{P3: New Constraint Generation } 
The prompt used for generating and curating new constraints is shown in Figure~\ref{fig:stage3_prompt}.

\begin{figure*}[h!]
\centering
\begin{tcolorbox}[title=Prompt 3: New Constraint Generation and Merging, colback=gray!5!white, colframe=black!75!white, fontupper=\ttfamily\small, width=\linewidth]
\textbf{Goal:} Generate additional high-quality constraints, merge them with extracted ones, and curate the final list for clarity, coverage, and non-redundancy. \\[2mm]

\textbf{Context:} You are given a programming \texttt{instruction}, a list of \texttt{relevant\_categories}, and the \texttt{extracted\_constraints} identified earlier. Your task is to expand and refine this list to produce the definitive set of constraints for this benchmark entry. \\[2mm]

\textbf{Process Overview:}
\begin{enumerate}[leftmargin=1.5em, itemsep=1pt, parsep=0pt, topsep=3pt]
    \item \textbf{Generate New Constraints:} Based on the instruction and relevant categories, produce 5--10 new constraints that are specific, objective, and verifiable. Tag each with \texttt{"instruction\_part": "Newly Generated"}.
    \item \textbf{Combine and Curate:} Merge newly generated and extracted constraints, ensuring coverage of all categories.
    \item \textbf{Refine and Resolve:} Remove duplicates, resolve contradictions, and prioritize clarity. Retain extracted constraints if overlap exists.
    \item \textbf{Finalize Count:} Ensure the final list has unique, valid constraints.
    \item \textbf{Validate Alignment:} Ensure every constraint’s type matches one of the provided \texttt{relevant\_categories}.
\end{enumerate}

\textbf{Principles for All Constraints:}
\begin{itemize}[leftmargin=1.5em, itemsep=1pt]
    \item Must be actionable, precise, and objective.
    \item Allow additional code when needed for compliance.
    \item No unresolved directive tokens (e.g., \{\{keyword\}\}).
\end{itemize}

\textbf{Output Format:}
\begin{lstlisting}[style=cifejson]
{
  "final_comprehensive_constraints": [
    {
      "type": "Constraint_Category_Name",
      "constraint": "Final, specific, objective, and atomic statement.",
    }
  ]
}
\end{lstlisting}

\textbf{Inputs Required:}
\begin{lstlisting}[style=cifejson]
{
  "instruction": {instruction},
  "relevant_categories": {relevant_categories_str},
  "extracted_constraints": {extracted_constraints_str}
}
\end{lstlisting}
\end{tcolorbox}
\caption{Prompt used in Stage 3: New Constraint Generation and Merging.}
\label{fig:stage3_prompt}
\end{figure*}

\paragraph{P4: Final Relevance Validation.}
The prompt used for final relevance filtering is shown in Figure~\ref{fig:stage4_prompt}.

\begin{figure*}[h!]
\centering
\begin{tcolorbox}[title=Prompt 4: Final Relevance Validation and Filtering,
                  colback=gray!5!white, colframe=black!75!white,
                  fontupper=\ttfamily\small, width=\linewidth]

\textbf{Goal:} Critically assess each constraint against the simplified instruction, justify the decision, and return only those constraints that are directly relevant. \\[2mm]

\textbf{Context:} You are given an \texttt{instruction} (core task) and a list of \texttt{final\_comprehensive\_constraints}. For every constraint, you must first write a reasoning statement and then make a binary relevance decision grounded in that reasoning. \\[2mm]

\textbf{Procedure:}
\begin{enumerate}[leftmargin=1.5em, itemsep=1pt, topsep=3pt, parsep=0pt]
  \item Understand the instruction’s core intent.
  \item For \emph{each} constraint:
    \begin{itemize}[leftmargin=1.25em, itemsep=1pt]
      \item Write a brief reasoning paragraph explaining whether and how it supports the instruction.
      \item Set \texttt{is\_relevant} to \texttt{true} or \texttt{false} based on that reasoning.
      \item Mark vague, off-topic, overly generic, or conflicting items as \texttt{false}.
      \item Include documentation-related items only if they directly support the core task.
    \end{itemize}
  \item Build a filtered list containing only relevant constraints.
  \item Provide a short summary explaining removed items.
\end{enumerate}

\textbf{Output Format:}
\begin{lstlisting}[style=cifejson]
{
  "evaluated_constraints": [
    {
      "constraint": "The original constraint text.",
      "reasoning": "Why it is (not) relevant to the instruction.",
      "is_relevant": true
    }
  ],
  "filtered_relevant_constraints": [
    {
      "type": "Constraint_Category_Name",
      "constraint": "Relevant constraint statement.",
      
    }
  ],
  "reasoning_for_removal": "Summary of why certain constraints were removed."
}
\end{lstlisting}

\textbf{Inputs Required:}
\begin{lstlisting}[style=cifejson]
{
  "instruction": {instruction},
  "final_comprehensive_constraints": {constraints_str}
}
\end{lstlisting}
\end{tcolorbox}
\caption{Prompt used in Stage 4: Final Relevance Validation and Filtering.}
\label{fig:stage4_prompt}
\end{figure*}

\noindent The combination of these four stages yields a constraint set that is aligned with developer intent while remaining clear, atomic, and evaluable.
% --- Appendix: Prompt — Constraint Quality Validation ------------------------
\paragraph{PQ: Constraint Quality Validation.}
The full prompt used for quality scoring (\textit{Atomicity}, \textit{Relevance}, and \textit{Objectivity}) is shown in Figure~\ref{fig:quality_prompt}.

\begin{figure*}[h!]
\centering
\begin{tcolorbox}[
  title={Prompt: Constraint Quality Validation (Atomicity, Relevance, Objectivity)},
  colback=gray!5!white,
  colframe=black!75!white,
  fontupper=\ttfamily\small,
  width=\linewidth ]

\textbf{System Role.} You are an expert in meticulously evaluating the quality of programming constraints.  
For each constraint, assign scores for \textit{Atomicity}, \textit{Relevance}, and \textit{Objectivity}, provide reasoning and improvement suggestions, and synthesize a unified quality score and overall analysis.  
Your judgment must be impartial and directly tied to the provided definitions.\\[2mm]

\textbf{Inputs Provided.}
\begin{lstlisting}[style=cifejson]
Original Instruction: {original_instruction}

Original Code (for context, if available):
```python
{original_code}

List of Generated Constraints to Evaluate:
{generated_constraint_list}
\end{lstlisting}

\textbf{Constraint Quality Criteria (Scores range from 1–5).}

\textbf{Atomicity (1–5):} Measures whether a constraint expresses exactly one indivisible requirement.

1 (Non-Atomic): "Return a float and raise ValueError for invalid input."

5 (Atomic): "Raise ValueError for invalid input."

\textbf{Relevance (1–5):} Measures how well the constraint aligns with the core task.

1 (Off-topic): "Avoid using global variables."

5 (Directly relevant): "Raise ValueError if the input DataFrame is empty."

\textbf{Objectivity (1–5):} Measures whether the constraint can be evaluated without subjective judgment.

1 (Subjective): "The code should be intuitive and clean."

5 (Objective): "The function must return a list of integers."

\textbf{Required Output Format.}
\begin{lstlisting}[style=cifejson]
{
"constraint_evaluations": [
{
"constraint_text": "The exact text of the constraint.",
"atomicity_score": int, // 1 to 5
"relevance_score": int, // 1 to 5
"objectivity_score": int, // 1 to 5
"reasoning": "Detailed explanation for each score, with suggestions for improvement."
}
// one entry per input constraint
],
"avg_atomicity": float,
"avg_relevance": float,
"avg_objectivity": float,
"unified_quality_score": float, // average of the three above
"overall_analysis": "Summary of overall quality, strengths, and weaknesses."
}
\end{lstlisting}

\end{tcolorbox}
\caption{Prompt used in the Constraint Quality Validation stage for assessing Atomicity, Relevance, and Objectivity of generated constraints.}
\label{fig:quality_prompt}
\end{figure*}

% --- Appendix: Instruction–Constraint Merging --------------------------------
\section{Merging Constraints into a Unified Instruction}
\label{app:merge-instruction}

This step produces a \emph{single, executable instruction} by weaving validated constraints back into the original task description without altering its semantics or structure. The merger preserves all original content (including punctuation, line breaks, code formatting, and any \texttt{BEGIN SOLUTION} blocks), inserts each constraint at a contextually natural location, and paraphrases minimally for fluency. The outcome is a unified prompt that downstream models can follow without juggling separate instruction and constraint lists.

\paragraph{Prompt.}
Figure~\ref{fig:merge_prompt} shows the prompt used to merge constraints into the instruction.

\begin{figure*}[h!]
\centering
\begin{tcolorbox}[
  title=Prompt: Merge Constraints into a Unified Instruction,
  colback=gray!5!white, colframe=black!75!white,
  fontupper=\ttfamily\small, width=\linewidth,
  enhanced, breakable, before upper=\raggedright
]
\textbf{Context:} I have an instruction for an LLM to generate a response and a set of conditions the model must follow while generating that response.

\textbf{Task:} Take the instruction and the conditions provided, insert the conditions into the instruction, and return the new instruction.

\textbf{Rules:}
1) For every condition, find a relevant position in the instruction where the condition can be inserted naturally.  
2) Paraphrase the condition only as needed so it fits fluently.  
3) Do \emph{not} delete, reorder, or alter any original content.  
4) Keep all punctuation, line breaks, return types, and code formatting intact.  
5) For solution snippets demarcated by \texttt{BEGIN SOLUTION}, you \textbf{must not} modify or remove them.

\textbf{Input format:}
\begin{lstlisting}[style=cifejson]
{
  "Instruction": {original_instruction},
  "Conditions": {new_constraints}
}
\end{lstlisting}

\textbf{Response format:}
\begin{lstlisting}[style=cifejson]
{
  "new_instruction": "The new instruction with the conditions merged in a natural way."
}
\end{lstlisting}
\end{tcolorbox}
\caption{Prompt used to merge constraints into a single, unified instruction.}
\label{fig:merge_prompt}
\end{figure*}

\begin{figure*}[h!]
\centering
\includegraphics[width=\linewidth]{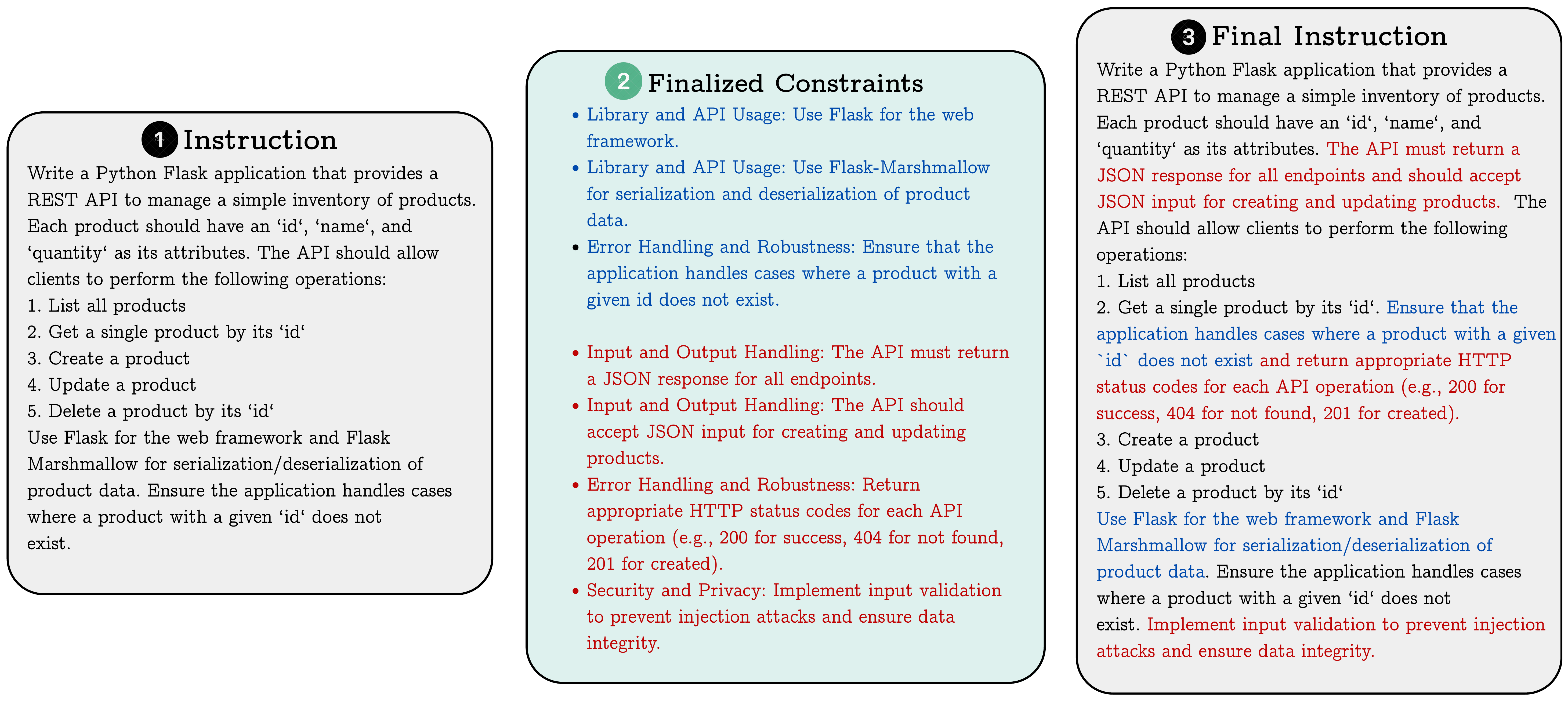}
\caption{Illustration of the instruction and constraint merging process. 
(1) The original \textbf{instruction} serves as the base text. 
(2) The \textbf{finalized constraints} are shown alongside where \textcolor{blue}{blue} highlights constraints \textit{extracted from the instruction} and \textcolor{red}{red} denotes \textit{newly generated constraints}. 
(3) These constraints are then seamlessly integrated into the instruction at contextually appropriate insertion points, resulting in the \textbf{final unified instruction}.}
\label{fig:merge_example}
\end{figure*}

\paragraph{Worked Example.}
Figure~\ref{fig:merge_example} illustrates the process: given the original instruction and the finalized constraints, the merger yields a unified instruction that preserves original wording and formatting while integrating each constraint at a semantically appropriate location.

% --- Appendix: Human Annotation ----------------------------------------------
\section{Human Annotation}
\label{app:human-annotation}

To ensure the reliability of LLM-based evaluations, we conducted a human annotation study focusing on two components of the benchmark: \textbf{constraint quality validation} and the \textbf{LLM-as-a-Judge evaluation} for constraint adherence. Two professional software developers with substantial experience in Python programming served as human annotators. Each annotator independently scored a randomly sampled subset of constraints along the three quality dimensions \textit{Atomicity}, \textit{Relevance}, and \textit{Objectivity}—as defined in the main paper.

The resulting inter-annotator agreement for constraint quality validation was \textbf{0.86}, indicating strong consistency in human judgment. When compared to the LLM-based quality assessment, the LLM–human agreement reached \textbf{0.84}, demonstrating close alignment between LLM-generated and expert-provided scores. Similarly, for the \textbf{LLM-as-a-Judge} evaluation of constraint adherence, the inter-annotator agreement was \textbf{0.84}, and the LLM–human agreement was \textbf{0.83}. These results collectively suggest that the proposed evaluation pipeline maintains high alignment with human expertise while offering the scalability benefits of automated assessment.
\paragraph{Agreement Metric.}
All agreement values reported in Table~\ref{tab:human_annotation} were computed using \textbf{Cohen’s $\kappa$ coefficient}, which measures the level of agreement between annotators while correcting for chance agreement. 
\begin{table*}[h!]
\centering
\begin{tabular}{lcc}
\hline
\textbf{Evaluation Component} & \textbf{Inter-Annotator} & \textbf{LLM–Human} \\
\hline
Constraint Quality Validation & 0.86 & 0.84 \\
Constraint Adherence (LLM-as-a-Judge) & 0.84 & 0.83 \\
\hline
\end{tabular}%

\caption{Agreement scores between human annotators and between human and LLM evaluations across benchmark components.}
\label{tab:human_annotation}
\end{table*}

\section{LLM-as-Judge Prompts}
We provide the exact prompts used to evaluate (i) constraint adherence (Figure~\ref{fig:llm_judge_constraint}) and (ii) functional correctness (Figure~\ref{fig:llm_judge_correctness}). Each prompt produces a structured JSON output to enable deterministic and reproducible evaluation.

% ---------------- Constraint Adherence ----------------
\begin{figure*}[h!]
\centering
\begin{tcolorbox}[title=Prompt: Constraint Adherence,
                  colback=gray!5!white, colframe=black!75!white,
                  fontupper=\ttfamily\small, left=1.2mm, right=1.2mm]
You are a verifier. Your task is to evaluate whether a given response satisfies a set of constraints for a specific instruction.

You will be provided:
- An instruction
- A list of constraints
- A response to the instruction (code block)

Your task:
- Analyze the response against each constraint independently.
- For each constraint, determine whether it is satisfied and give a brief explanation.
- Do not assume facts not present in the response or the constraints.

\textbf{Required JSON output (and only this JSON):}
\begin{verbatim}
{
  "Evaluation": [
    {
      "Constraint": "<constraint text>",
      "Reason": "<why satisfied / violated>",
      "Aligns": true | false
    }
    // ... one object per constraint
  ]
  }
  \end{verbatim}

\end{tcolorbox}
\caption{LLM-as-Judge prompt for constraint adherence.}
\label{fig:llm_judge_constraint}
\end{figure*}

% ---------------- Functional Correctness ----------------
\begin{figure*}[h!]
\centering
\begin{tcolorbox}[title=Prompt: Functional Correctness,
                  colback=gray!5!white,
                  colframe=black!75!white,
                  fontupper=\ttfamily\small,
                  left=1.2mm,
                  right=1.2mm]
You are an expert Python developer and code reviewer.  
Your task is to evaluate whether a given Python code correctly follows the provided instruction.

\textbf{Evaluation Criteria:}
\begin{itemize}[leftmargin=1.2em, itemsep=1pt]
  \item \textbf{Completely Correct:} The code fully satisfies the instruction and contains no syntax or semantic errors.
  \item \textbf{Partially Correct:} The code mostly satisfies the instruction and is syntactically and semantically valid but may miss some edge cases or minor implementation details.
  \item \textbf{Wrong:} The code contains syntax or semantic errors, or clearly fails to follow the instruction.
\end{itemize}

\textbf{Output Format:}
Return the evaluation strictly as a JSON-style dictionary without any additional explanation:
\begin{verbatim}
{
  "reason": "<Your reason for the evaluation>",
  "correctness": "Completely Correct / Partially Correct / Wrong"
}
\end{verbatim}

\textbf{Input Format:}
\begin{verbatim}
Instruction:
<instruction>

Generated Code:
```python
<generated_code>
\end{verbatim}
\end{tcolorbox}
\caption{LLM-as-Judge prompt for evaluating functional correctness of generated code.}
\label{fig:llm_judge_correctness}
\end{figure*}

\section{Results}
\begin{figure*}[t]
    \centering
    \includegraphics[width=\linewidth]{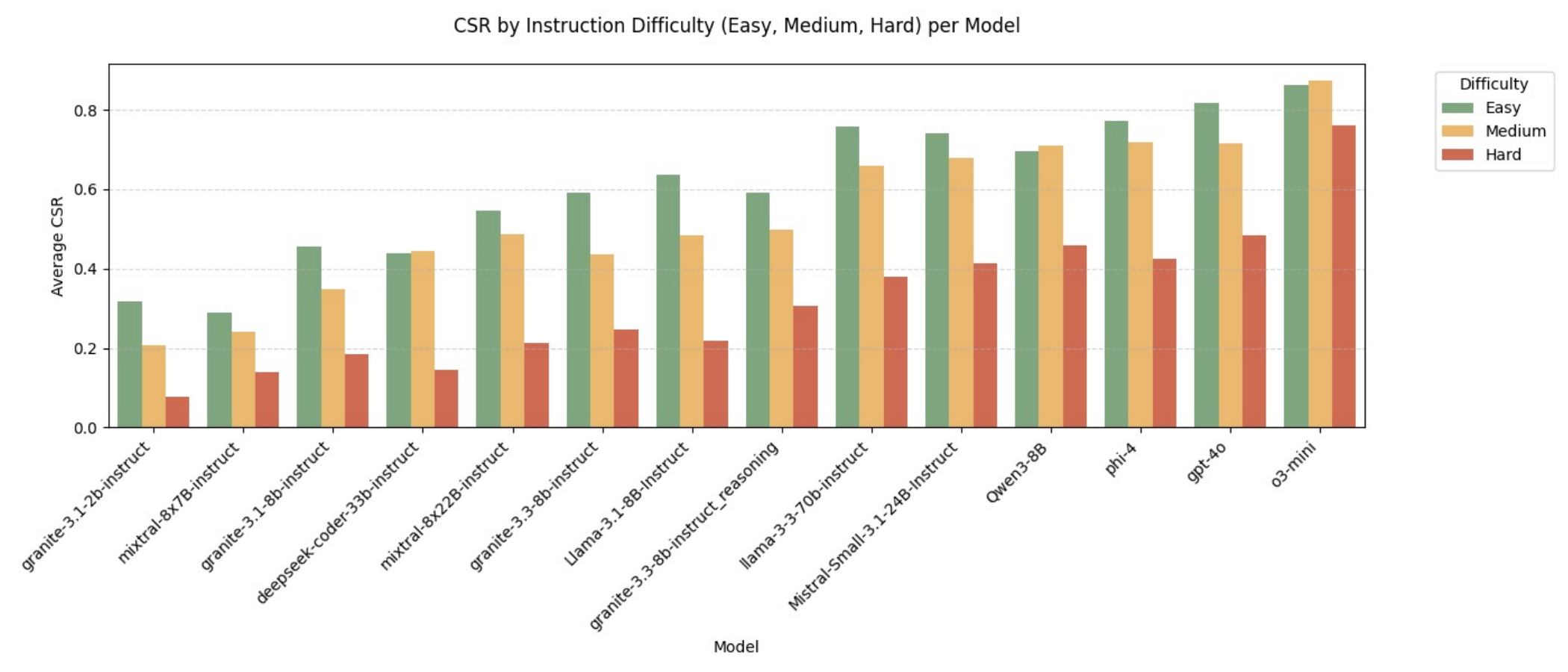}
    \caption{CSR by instruction difficulty (easy, medium, hard). All models degrade with harder tasks, though reasoning-enhanced models show more robust performance.}
    \label{fig:csr_by_difficulty}
\end{figure*}

\bigskip